# Learning Organization using Conversational Social Network for Social Customer Relationship Management Effort

Andry Alamsyah,[1,2] Yahya Peranginangin,[1] Gabriel Nurhadi,[1]

[1]*School of Economics and Business, Telkom University, Indonesia*
*E-mail: andrya@telkomuniversity.ac.id*
[2]*School of Electrical Engineering and Informatics, Institut Teknologi Bandung, Indonesia*

The challenge of each organisation is how they adapt to the shift of more complex technology such as mobile, big data, interconnected world and Internet of things. In order to achieve their objective, they must understand how to take advantage of interconnected individual inside and outside the organisation. Learning organisation continues to transform by listening and maintain connection with their counterparts. Customer relationship management is an important source for business organisations to grow and to assure their future. The complex social network, where interconnected peoples get information and get influenced very quickly, certainly a big challenge for business organisations. The combination of these complex technologies provides intriguing insight such as the capabilities to listen what the markets want, to understand their market competition, and to understand their market segmentation. In this paper, as a part of organisation transformation, we show how a business organisation mine online conversational in twitter related to their brand issue and analyse them in the context of customer relationship management to extract several insight regarding their market.

1. Introduction

In every organization, there are challenges on how they adapt to their environment in order to sustain their survival or to increase their influence. Continuously transformed organizations are desirable, but yet it is a difficult and complex process to achieve as an organization becomes larger. One of the examples is organization technology adoption, the famous Bell's technology adoption lifecycle describe most of individuals will be conservative to new technology. This fact contradicts to the urgently needed rapid organisational transformations in order to adapt and to compete. Learning organization proposed a framework on how an organization think and act in more complex and interconnected way [1]. This framework provides opportunity on how an organization transform in smart way.

The arrival of social media provides a conversation platform that stimulates the generation of data with characteristics such as large volume, fast streaming, real-time and rich variety. These characteristics lead to the term of Big Data. Big Data can help organization to have thorough knowledge by listening to the crowd of social media and Internet users [2]. Knowledge management is enriched by feeds of these kinds of data [3][4]. Having advanced knowledge will certainly help to speed up organizations transformation to be more adaptive to their environment. Customer relationship management will also improve significantly, because the decision maker have enough knowledge to act to customer needs dynamically and even personally.

Business organization important tasks are to listen and to know what market wants and what their current competitions are. The large majority of data collection effort is using offline approach directly from population using methods such as questionnaire, interview, snowball sampling, contact tracing, random walk and direct informations. These methods are practically very good if we deal with small-medium number of data. In the case of larger crowd such as conversations in social media those approaches are very expensive, time consuming process and having the accuracy issue. These shortcomings are typical to offline data collection approach, where they get complicated when population is getting bigger.

The objective of this paper is to show how we can use social network based on online conversation to understand the dynamics of the network/market such as the influential actor in the network, market segmentation, how information flow in the network, and some others. These informations are important for business organisations to react agile to the dynamics of the market. We conduct the experiment using social network conversation in social media Twitter. Those conversations are modeled as social network models based on graph theory. Graph theory provides some properties such as density, diameter, trees, connectivity, shortest-path, node degree and some others [5]. Those properties can be expanded to measure network topology, which will be explained in next chapter.

2. Social Networks Models for Complex Relationship

To understand pattern and behavior of social networking, they are modeled based on graph theory. *Graph theory* is a branch of mathematics combinatorics where we represent entities as vertices and relationship among the entities as edges [6]. Once we have network, which consists of vertices and edges, we will have the properties such as the density of global network or partial network, network diameter, the giant components, clustering coefficient, number of communities and other useful information. Those properties are important as metrics and quantitative measures of the social network. Complex relationship is based on *Complex Network* [7], which defined as network with non-trivial topological features that do not occur in simple network, but often occur in real-world network. Social network is one example of complex networks. Complex relationship describe real-world social relationship, which rarely simple and straightforward to predict, involving human emotion and the influenced by human close-ties.

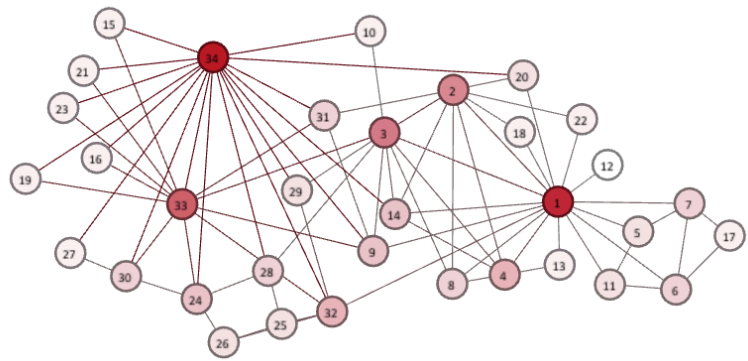

*figure 1. Network model of interaction between employees in knowledge sharing context*

The basic of social network analysis describe as the case study from knowledge transfer inside an organization of 34 employees regarding IT subject in figure 1 [8]. The vertices represent individual and the edges represent interaction, in this case, it is the transfer knowledge between individual. The redder the colour signifies the more interaction an individual has. From this knowledge network we found key relationship such as [8]:
    a. It is very important to know what someone knows; this is related to the ability to analyze knowledge and skills of organization members and also evaluate the overall cohesion of the network.
    b. We can gain fast access to certain people in the network; this is related to the ability to identify the most central/influential members of the organization.
    c. We can create workable knowledge through cognitive engagement; this is related to assess those who are not well connected in the network, these people are probably representing the underutilized assets.

d. There are learning processes in trust relationship; this is related to analyzing the network by highlight ties between people who we will trust in knowledge sharing information.

From the case study above we learn that we can better understand the dynamics of network formation and hopefully be able to intervene at the times of knowledge creation and sharing. The four keys relationship can be viewed separately as different aspect of knowledge management inside organizational network and also can be view accumulatively. In short, we have the answer of the question on who should we turn to ask for knowledge or information among our co-worker regarding IT subject. From the figure 1, we see that employee number 1, 33 and 34 are the most connected peoples, which means that they are the most likely peoples turn for advice. We also found a subgroup contains of six employees on the right side of the network, this subgroup will disconnect from the network if employee number 1 is not present to the network. The existence of this subgroup signifies inefficiency in knowledge utilization where member of the subgroup are not maximizing expertise utilization from the network. The strategy of strengthening out ties between individuals is important solution to broader the information access to the rest of the network.

The study of complex social relationship described above is called *Social Network Analysis* (SNA). SNA provides network quantification based on graph theory. Some often-used metrics in SNA are centrality, community detection, homophilly, reciprocity, structural holes, bridge, overlapping communities, component, connectivity[6] [9] [10]. The metrics are important indications on how good or bad our network is and it can be used to as comparison tools between several different social networks. From the description above, it is clear that our research depends on network formation or topology and it does not depend on the content of communication. The workaround strategy for this approach is by defining context of conversation in the first place and then analyze network topology formed by specific context.

3. **Customer Relationship Management**

In General, *Customer Relationship Management* (CRM) is a strategy to oversee the customer activities [11]. It has been the strategic approach that most companies had taken in trying to figure out how to supervise their customer behavior. The methodology and technology used in CRM are destined to increase customer repeat purchase. Gathering data about the customer and tracking all customer activities were the way that CRM was used to find out individual customer's thinking. The positive customer experiences are mostly based on operational response. The feedbacks from customer are processed into CRM system to create better customer strategy. The objectives of CRM are to maximize customer value, to increase company profitability and shareholder value [12]. Traditionally, there are three main parts of CRM: Operational CRM for customer automation process, Collaborative CRM for providing customer communication channels, Analytical CRM for producing customer behavior information based on data mining algorithm and process. In SCRM, those functions are taking advantage of communities in social network and become social sales and social marketing. The difference between the two approaches is shown at figure 2.

The exponential growth of social media / Web 2.0 and the shifted of social communications to more on customer controls are leading to the birth of new strategy called *Social Customer Relationship Management* (SCRM) [13]. SCRM designed to engage the customer in a collaborative conversation in order to provide mutually beneficial value in trusted and transparent business environment. It is the company response to the customer's ownership of the conversation. One of SCRM strategies that we propose in this paper employs *Social Computing* approaches [14], which capture online conversation to understand users actions, user preferences, users behaviours and also the overall implications for the network. By using SCRM, the conventional CRM have developing broader than just a media to retain current customer. The presence in social network services has positive recognition from potential customers. One of the advantages of this system is considered important for brand awareness effort.

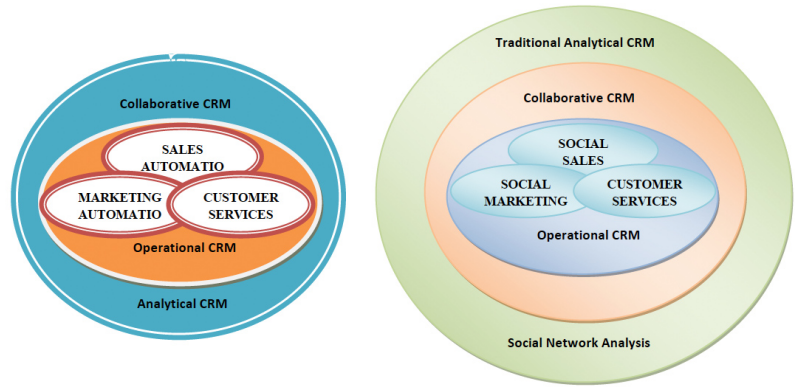

*figure 2. The difference approach betweeen CRM (left) and SCRM (right)* [11]

The SNA and *User Generated Content* (UGC) together with data and user profiles are four factors that need to be captured to help gaining access to customer insight. SCRM tools provide the means to capture data, the profiles, and to create the experience maps which in turn help develop the real insight into customers that provide what is a genuinely personalized and delineable experience for individual customers. Those actions were not supported by traditional CRM. Overall, the comparative features or functions between CRM and SCRM method shown in table 1. [11]

*Table 1 the comparative features or functions between CRM and SCRM method [11]*

| CRM | SCRM |
|---|---|
| Definition: CRM is a philosophy and a business strategy, supported by a system and a technology, designed to improve human interactions in a business environment. | Definition: Social CRM is a philosophy and a business strategy, supported by a system and a technology, designed to engage the customer in a collaborative interaction that provides mutually beneficial value in a trusted & transparent business environment. |
| Tactical and operational: Customer strategy is part of corporate strategy. | Strategic: Customer strategy is corporate strategy. |
| Relationship between the company and the customer was seen as enterprise managing customer - parent to child to a large extent. | Relationship between the company and the customer are seen as a collaborative effort. And yet, the company must still be an enterprise in all other aspects. |
| Focus on company not on customer relationship | Focus on all iterations of the relationships (among company, business partners, customers) and specifically focus on identifying, engaging and enabling the "influential" nodes |
| The company seeks to lead and shape customer opinions about products, services, and the company-customer relationship. | The customer is seen as a partner from the beginning in the development and improvement of products, services, and the company-customer relationship. |
| Business focus on products and services that satisfy customers | Business focus on environments & experiences that engage customer. |
| Customer facing features: sales, marketing and support. | Customer facing both features and the people who is in charge of developing and delivering those features. |
| Marketing focused on processes that sent improved, targeted, highly specific corporate messages to customer. | Marketing focused on building relationship with customer - engaging customer in activity and discussion, observing and re-directing conversations and activities among customers. |
| Intellectual property protected with all legal might available. | Intellectual property created and owned together with the customer, partner, supplier, and problem solver. |
| Insights and effectiveness were optimally achieved by the single view of the customer (data) across all channels by those who needed to know. Based on "complete" customer record and data integration. | Insights are a considerably more dynamic issue and are based on customer data, customer personal profiles on the web and the social characteristics associated with them, customer participation in the activity acquisition of those insights. |
| Resided in a customer-focused business ecosystem. | Resides in a customer ecosystem |

| Tools are associated with automating Functions. | Integrates social media tools such as blogs, wikis, podcasts, social networking tools, content sharing tools, user communities into apps/services |
|---|---|

## 4. Experiment and Analysis

To show SCRM implementation, we conduct an experiment using conversation data from social media Twitter. We crawl the conversation through Twitter Application Programming Interface (API). This API offers simpler way of gathering data by providing uniform format. Twitter API is much more efficient than using technique such as web mining, which proved difficult regarding variety formats and data types available on the web. Another reason is the practicality of Twitter as open platform services and one of the most active social network services for online conversation in Indonesia. Our research workflow is shown at figure 3.

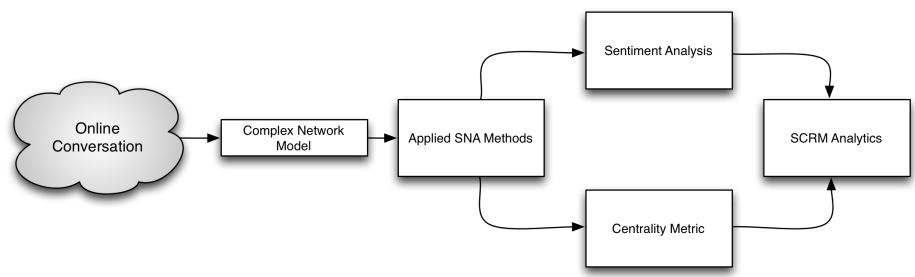

*figure 3. The research workflow*

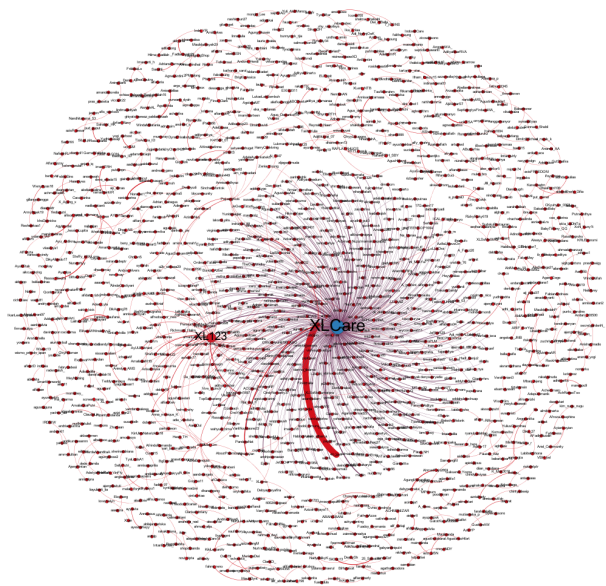

*figure 4. The complex social network conversation model*

For the experimentation, we choose one of big telecommunication company in Indonesia, which is PT. XL Axiata Tbk. Based on our hypotheses and observations, Indonesian telecommunication subscribers has tendency to post their opinion or complain about unsatisfactory service they have received on social media, including Twitter. The Twitter data crawling is set for period July 1-3th, 2013 for all tweets that mention keyword and hashtag *"XL"*, *"XLCare"*, *"XL123"*. We get undirected network with 1413 peoples involved in the conversation and 1026 conversation between those peoples. Thus, we get 1413 nodes and 1026 edges. The result of the whole network graph model is in figure 4.

We apply centrality metrics to find the most influential actors in the network. In this paper, we focus on three centralities metric: degree, betweenness and closeness. The degree centrality (DC) is to measure total number of connections or the most connected actor in the network. Once we know DC value, we can identify which actors in the conversation network that actively connected to many other actors, probably they are the one who respond to many questions from the crowd. The betweenness centrality (BC) is to measure the importance of actor location in the network. BC value will show which actors are the hub or bridge between different part of communities / group in the network. The hub actors is important to fill any structural hole in the network, their presence are strengthening the network structure. The closeness centrality (CC) is to measure how fast an actor reaches all other actors in the whole network. By knowing the CC value, we can assign set of actors who will spread the information fastest to the whole network. Our centrality metric measurement rank and value is shown at table 2.

Table. 2. Degree centrality, betweenness centrality and closeness centrality metric value and rank

| Ranks | Node | Degree Centrality | Node | Betweenness Centrality | Node | Closeness Centrality |
|---|---|---|---|---|---|---|
| 1 | XLCare | 437 | XLCare | 0.130 | XLCare | 0.768 |
| 2 | XL123 | 59 | XL123 | 0.020 | XLSoMeSumatera | 0.750 |
| 3 | XLandMe | 16 | XLandMe | 0.006 | parkiyeeon | 0.750 |
| 4 | PejuangKuis | 13 | PejuangKuis | 0.003 | xxopeter | 0.750 |
| 5 | Viccent22 | 6 | AfinaTsabbita | 0.003 | MomeafmPLM | 0.750 |
| 6 | RAFLATAHUGS | 5 | Viccent22 | 0.003 | azizanangM | 0.667 |
| 7 | TanteYulia | 4 | Serbakuis | 0.002 | yonkya | 0.667 |
| 8 | Adhantriio | 4 | adhitstubz | 0.002 | BeautyCare19 | 0.667 |
| 9 | Widideon | 3 | adiUFO | 0.002 | bandreanto | 0.667 |
| 10 | Zulfincitra | 3 | anisha_ir | 0.002 | Exploso | 0.667 |

From the table 2., we see that @XLCare is consistently at the top of all centrality measurement. Since this account is the official account of the company, the result is not surprising us. In order to better understand the community / social / crowd contribution, we remove the official account from the end result. We found @PejuangKuis is non-official account, while its important actor, it does not sufficiently generate and involve in the conversation. This is a warning for the company that their network does not form closed-ties, the role of crowd is minimal, and they rely heavily on official account. In the future, they need to implement a strategy to maximize the role of the crowd in support brand or customer relationship management activity.

When we look into content of the conversation, we got 745 conversation (52.72%) contains of negative sentiment, which mainly bad opinion and complaints to customer services. We got only 102 conversations (7.22%) positive sentiment. When we look at this composition, the tendency in today online social media people speak up when they have bad experience or they are actually having trouble with the services. Our ability to implement effective and efficient SCRM strategy is important in controlling the content or the structure of the network. Using the knowledge of SNA and all the metrics available will certainly help us with the effort to form the network as we intend to.

5. **Conclusion**

This paper has shown a workflow of a SCRM effort though Twitter. The SCRM effort can be modeled using complex social network. To guarantee the effectiveness of SCRM effort we need to implement a strategy with objective to increase closed-ties between actors in network, which lead to self-healing management network. The key of our preferred network found on positive sentiment. From our observations, the strategies such as thematic effort, contextual storytelling, and appreciation to customer can lead to positive vibes in the network. For the future research, there is a need to find the academic connection between network with positive sentiment with willingness of the crowd to help the communities or willingness to defend the brand / organization.

This study can be enhance, by extending the length of observations to get larger data set / conversation for more accurate result. Another path that might be interesting to follow is to combine several social network services crawling data for more complete insight about what actually happen in the market. The metric combinations will also certainly useful to identify individual behavior in the network.

SCRM implementation is one way to apply smart learning in business organization facing the more interconnected world. Data and tools are widely available, and it is up to the top management whether they are going to adapt the new way to get closer to the customer. Most of organizations will have trouble to embrace this approach because of the lack of sufficient people who understand working with stream and large quantity of data.

**Acknowledgement:**

NoLimitID who provide us with Indonesian Twitter data conversations to supports this work.

**References:**

[1]. Senge. P.M. *The Art and Practice of The Learning Organization: The New Paradigm in Business: Emerging Strategies for Leadership and Organizational Change.* 2$^{nd}$ Edition. London; McGraw-Hill. 1990
[2]. MacAfee. A, Brynjolfsson. E. "*Big Data: The Management Revolution*". Harvard Business Review Magazine. October 2012
[3]. Hislop. D. "*Knowledge Management in Organizations*". Oxford University Press. 2005
[4]. Alamsyah. A. "*Role of Social Network Analysis in Knowledge Management*". Jurnal Manajemen Indonesia Vol 12 No 4. Pp 309-314. 2013
[5]. Diestel. R. "*Graph Theory: Electronic Edition 2005*". Springer-Verlag Heidelberg, New York, 1997, 2000, 2005
[6]. Scott. J. "*Social Network Analysis Theiry and Applications*". Sage Publications. 2000
[7]. Cohen. R, Havlin. S. *Complex Network : Structure, Robustness and Function.* Cambridge University Press. 2010
[8]. Alamsyah. A, Peranginangin. Y. "*Effective Knowledge Management using Big Data and Social Network Analysis*". International Journal of Learning Organization: Management and Business Vol 1 No 1. 2013
[9]. Newman.M.J. "*Network: An Introduction*". University of Michigan and Santa Fe Institute. Oxford University Press. 2010
[10]. Alamsyah. A, Rahardjo. B, Kuspriyanto. "*Social Network Analysis Taxonomy Based on Graph Representation*". Proceeding of The 5$^{th}$ Indonesian International Conference on Innovation, Entrepreneurship, and Small Business. 2013
[11]. Mosadegh. M.J., Behboudi. M. "*Using Social Network Paradigm for Developing a Conceptual Framework in CRM*". Australian Journal of Business and Management Research. Vol 1, No 4, pp 63-71. 2011
[12]. Payne.A, Holt. S, Frow.P. "*Integrating Employee Customer and Shareholder Value Through an Enterprise Performance Model: an Opportunity for Financial Services*". International Journal of Bank Marketing. 18(6), pp 258-273. 2000

[13]. Greenberg. P. *"The Impact of CRM 2.0 on Customer Insight"*. Journal of Business and Industrial Marketing. Vol. 25 Issue 6. pp 410-419. 2010

[14]. Wang. F, Zeng. D. *"Social Computing From Social Informatics to Social Intelligence"*. Journal of Intelligent System IEEE. 2007